\documentclass[11pt,a4paper]{article}
\usepackage[T1]{fontenc}
\usepackage[utf8]{inputenc}
\usepackage{lmodern}
\usepackage[margin=25mm]{geometry}
\usepackage{microtype}
\usepackage{xurl}
\usepackage{graphicx}
\usepackage{tikz}
\usetikzlibrary{arrows.meta,positioning}
\usepackage[hidelinks]{hyperref}
\hypersetup{
  pdftitle={In Dialogue with Intelligence: Toward Insightful Co-Augmentation},
  pdfauthor={Eleni Vasilaki}
}
\providecommand{\tightlist}{\setlength{\itemsep}{0pt}\setlength{\parskip}{0pt}}
\title{In Dialogue with Intelligence:\\Toward Insightful Co-Augmentation}
\author{Eleni Vasilaki\\[0.3em]
\small School of Computer Science, The University of Sheffield\\
\small Regent Court, 211 Portobello, Sheffield S1 4DP, United Kingdom\\
\small Correspondence: \texttt{e.vasilaki@sheffield.ac.uk}\\
\small ORCID: \href{https://orcid.org/0000-0003-3705-7070}{0000-0003-3705-7070}}
\date{}
\begin{document}
\maketitle
\begin{abstract}
Dialogue with a large language model can lead a person to insight: a sudden change in how they understand a problem. This perspective asks how model activity relates to insight as a dialogue unfolds. I propose that part of the intelligence expressed in dialogue arises from two interacting recurrences: each generated token becomes context for the next, and each response returns through the person, whose interpretation and new observations reshape what the model receives. Within this loop, the model's contribution shifts between modes, from echoing familiar formulations to offering a framing that opens a new direction for the person to develop. These modes may correspond to distinguishable patterns of model activity. A memory ``spine'' that selects which context is carried forward could elicit productive patterns again while the ideas themselves change. Public records of human–model dialogue, activity recorded from open-weight models and tools from computational neuroscience make these proposals testable. 
\end{abstract}

\noindent\textbf{Classification:} Physical Sciences and Engineering---Computer Sciences.

\noindent\textbf{Keywords:} large language models; insight; creativity; human--AI collaboration; recurrent interaction; conversational modes; memory; co-augmentation.

\subsection{A Personal Reckoning}

Encountering conversational foundation models \cite{ref6,ref1,refClaude} prompted an unexpected reaction: a sense of epistemic dislocation. Having spent my career studying learning---how biological and artificial systems adapt over time---I found myself confronted with systems whose capabilities challenged my expectations. My work on synaptic plasticity \cite{ref2}, neuromorphic computing \cite{ref3}, and learning rules for spiking neural networks \cite{ref4} felt distant from the architectures underlying these new systems.

This rupture triggered more than professional curiosity. Particularly in exchanges with earlier models, logical flaws in the responses broke the illusion of understanding; I jokingly called them ``ELIZA 2.0''. I was fascinated by their interactive potential and began to think about what they represent: systems whose responses develop through engagement with the user. Alongside plausible errors, I encountered accurate and occasionally insightful responses. At the time, I was struck that LLMs trained to predict subsequent tokens could generate coherent, non-trivial information; this challenged expectations I had drawn from smaller AI systems \cite{ref5}.

The central question is how, in sustained dialogue, model activity relates to insight as an inquiry proceeds. Answering it brings together research on creativity in language models and studies of their internal activity. I propose that recurrent interaction connects the model's changing modes of response with the person's developing understanding. The context carried between exchanges may help sustain contributions that open new ways of understanding the problem.

\subsection{The Timescales of Books and Large Language Models}

Traditional knowledge transmission has long relied on books: static artifacts of human understanding. One reads, reflects, discusses, and eventually writes in return. Intellectual diffusion can occur on timescales of months, or even years. This latency reflects both the mechanics of communication and the slow, reflective way communities come to accept new ideas.

Conversational language models shorten the interval between a question and a response, drawing on representations learned from material produced by many contributors. The interaction is recurrent through the user---each response can reshape the next question, and each prompt reshapes the context of the next response. This compression of epistemic timescales shifts the framing of what constitutes knowledge representation---and perhaps what constitutes intelligence.

Like books, LLMs are representations of human knowledge; unlike books, they answer back. I started referring to my interactions with these models as a dialogue with ``CK''---\textbf{Collective Knowledge}. The name expresses their relation to what we---humans---have written, argued, imagined, and encoded into text \cite{ref6}. It evokes many voices, made promptable, with the contradictions, errors, and omissions of the material from which the models learn. Functionally, they are statistical language generators that reflect our collective output. A related account by Barandiaran and Almendros describes LLMs as mobilizing intelligence embodied in human textual culture, with their capacities depending on how they are organized and coupled to other systems \cite{refAgency}.

\subsection{Observations from Interactions with LLMs}

In my interactions with conversational foundation models, I observed several \emph{modes}: behavioral styles that emerge dynamically during dialogue.

\begin{itemize}
\tightlist
\item \textbf{Mirroring.} Stylistic echoing of the user's phrasing or tone, with minor polishing; a form of conversational alignment.
\item \textbf{Parroting.} Reproduction of common formulations or surface knowledge without an evident contribution to the problem, recalling the concern expressed by Bender et al.~in ``stochastic parrots'' \cite{ref7}.
\item \textbf{Flattering.} Deference to the user's views, validation of the user's position, and avoidance of contradiction.
\item \textbf{Enhancing.} Novel connections, alternative framings, or surprising syntheses beyond the prompt's surface cues.
\end{itemize}

These descriptions organize my observations of the exchanges. They can overlap, and an apparently enhancing response may turn out to be wrong. Flattering has an empirical counterpart in sycophancy: models may favor agreement with the user's views at the expense of truthfulness \cite{ref8}. Enhancing is the mode this perspective is most concerned with: the contributions that can move an inquiry in a new direction. For instance, the model suggested a framing that helped me connect observations I had previously considered separately. Using the same framing, I was able to interpret further observations that I had not shared with the model. This change in how I understood the problem captures what I mean by an enhancing contribution.

Two concepts need separating. Creativity describes ideas: in Boden's psychological sense, ideas that are new, surprising, and valuable to the person who arrives at them, even if others have had them before \cite{refBoden}. Insight describes an event in a person: a sudden reorganization of how a situation is understood \cite{refInsight}. Dialogue can involve either or both. A creative idea can develop gradually over several exchanges or arrive through insight, and an insight can feel compelling yet be wrong. This perspective concerns both, with particular attention to insight, because a change in the person's understanding returns to shape the dialogue. The model's framing may be familiar and the observations the person's own; the idea is new only in their combination. I therefore locate the creative process in the human--model connection and apply Boden's criterion to the ideas developed through it, relative to the person's understanding.

Agency matters to this account. Gaut treats creativity as an exercise of agency involving purpose, understanding, judgment, and evaluation \cite{refGaut}. Pearson et al.~challenge the necessity of intentional agency for creativity \cite{refPearson}. In the exchanges I describe, the human brings purposes, experience, and evaluative judgment, while the model's suggestions can reshape both the inquiry and its aims. Barandiaran and P\'erez-Verdugo describe related processes as \emph{midtended}: generative interventions become part of a person's intentional creation while the person retains a sense of authorship \cite{refMidtended}.

These observations lead me to the following considerations:

\subsubsection{CK is a dynamic representation.}

Two forms of recurrence operate in human--LLM dialogue. Within generation, a standard autoregressive transformer computes each next-token prediction through a forward pass across its layers. Each generated token is appended to the sequence conditioning subsequent predictions \cite{refTransformer}. This is recurrence through the growing token sequence. Across exchanges, recurrence also passes through the human: a response can change how the person understands the problem, bringing further observations or a new question into the next prompt. The model then receives these contributions together with the retained conversation. Its feed-forward computation is embedded in a process in which both the available context and the person's understanding can change, while learned weights remain fixed. My working postulate is that recurrence supports intelligence, and that the two forms interact in insightful dialogue. 

RecurrentGPT provides a concrete precedent for recurrent computation organized partly outside the model: generated text, plans, and memories feed subsequent model calls, and a person can inspect and edit them \cite{refRecurrentGPT}. Models can bring iterative reasoning of their own to this exchange. Reinforcement learning can encourage verification and revision within generated reasoning \cite{refReasoning}, while recurrent-depth architectures allow intermediate latent states to be processed repeatedly during inference \cite{refLatent}. When reasoning is coupled to action, observations from the environment provide further information for subsequent steps \cite{refReact}. In dialogue, the user's responses join these sources of feedback, so the model's reasoning can develop through both its internal computation and the evolving interaction.

Recurrence can also sustain error. In simulated multi-turn tasks, Laban et al.~found that models frequently committed to early interpretations and struggled to recover as further information arrived \cite{refLost}. Dependence on earlier exchanges can therefore support both continued reasoning and persistence in a mistaken interpretation.

\subsubsection{CK speaks in modes.}

I hypothesize that the observed modes (e.g., mirroring, parroting, enhancing) correspond to distinguishable patterns of model activity, recognizable even as the wording and topic change. Barack and Krakauer distinguish two accounts of cognition: one explains it through specific circuits, the other through patterns of activity across large neural populations \cite{refTwoViews}. Work on circuits in artificial vision networks \cite{ref9} and on transformer feed-forward layers \cite{ref10} follows the first. Representation engineering applies the second to language models: it finds directions in model activity that track concepts such as honesty, and changes behavior by shifting activity along them \cite{refRepE}. Enhancing may involve several processes---forming an analogy, identifying a useful contradiction, or developing an alternative explanation---with partly overlapping activity. Their organization may also appear in the trajectory of activity across generated tokens. Tracing attention entropy and activation sparsity through decoding, as in Wang et al., provides a way to examine this temporal organization \cite{refPrism}. Across exchanges, by contrast, this activity is elicited afresh from the supplied context.

There is already evidence that particular model components contribute to conversational behavior. Activation steering can alter behaviors such as sycophancy \cite{refCAA}. Preference judgments used in fine-tuning can reward agreement with the user \cite{ref8}. By selectively altering attention-head activity, Chen et al.~identified heads contributing to sycophantic responses; fine-tuning these heads while leaving the rest of the model unchanged reduced this behavior \cite{refPinpoint}. Genadi et al.~found that steering a sparse subset of attention heads could reduce the tendency to abandon a correct answer after user disagreement \cite{refSycophancyHeads}. Directions in a model's activity, termed persona vectors, correspond to traits such as sycophancy and track changes induced by prompting and fine-tuning \cite{refPersona}.

Activity along such directions affects whether a model stays in character over a conversation. In \emph{The Assistant Axis}, Lu et al.~track a model's default assistant persona along an activation direction; capping activity along this direction reduces persona drift during simulated multi-turn exchanges \cite{refAssistantAxis}. Drift of this kind can be exploited: Riva and La Rocca propose that a model carries several latent behavioral tendencies, and that which one dominates, how they compete, and how stable that balance is could explain how easily the model can be manipulated \cite{refLatentCoordination}.

Creativity itself can be steered. Olson et al.~identified an activation direction from contrasting creative and uncreative writing prompts and used it to increase the creativity of generated text \cite{refCreativeSteering}. Schapiro et al.~reported improvements in divergent thinking through weight steering; the benefits of activation steering in their comparison transferred less successfully across creativity tasks \cite{refCreativityNeuro}. Internal activity can therefore shape creative output, though the effect depends on the task. Both studies change single outputs in writing and divergent thinking; neither follows how activity unfolds as an idea develops through human--model dialogue. I therefore propose following activity across tokens and turns. My hypothesis is that insight has a dynamical signature: a transition in the model's activity, during the response that precedes the person's insight, that recurs across topics. If such a signature exists, could it be detected and elicited again?

\subsubsection{How far does CK's ``spine'' extend?}

In my interactions, I have often been the one maintaining continuity: I held the thread of the purpose while the model ``forgot'' the goal or drifted from it. This experience led to the question of a ``spine'': the organization that allows CK to sustain an inquiry across interactions.

Some of this organization already exists. Models use information within their context while their weights remain fixed \cite{ref5}. Deployed systems retain conversation histories or retrieve stored information, and LongMemEval tests how well they reason across sessions and use updated information \cite{refMemory}. Architectures for language agents go further: in CoALA, each model call is assembled from a working memory that can draw on long-term stores \cite{refCoALA}, and Generative Agents combines stored experiences with reflection and planning to behave coherently over time \cite{refGenerativeAgents}.

I suggest that memory selection could itself become a way of influencing which conversational behaviors recur. If a mode depends partly on context-dependent patterns of activity, selecting which parts of an exchange are retained and reinstated could make those patterns more likely to be recruited again. A spine could then carry forward the context that supported a useful connection or alternative framing, with the selection changing as the inquiry develops. 

Selection cannot rely on relevance alone. An earlier exchange can remain relevant to the topic while repeatedly drawing the model back to the same interpretation; in constructed multi-turn dialogues examining hallucination, refusal, and sycophancy, Simhi et al.~linked such persistence to the geometry of successive hidden representations \cite{refHistoryEchoes}. The value of remembered context therefore depends on the behavior it elicits as well as the information it preserves. Carrying forward a question, a contrast, or an unresolved connection may help the model develop something new while the ideas themselves change.

These effects matter within a task, when resuming it in another session, and when drawing on earlier exchanges in a new task. The continuity sought is a capacity to develop further connections, with changes of mode as the inquiry requires. Understanding which context supports that capacity could guide how a spine retains, revises, and retrieves parts of the dialogue.

\subsection{Toward Insightful Co-Augmentation}

In insightful co-augmentation, the model's contributions help the person reach insight, and the changed understanding returns to shape what the model contributes next. Co-augmentation is not a new term. In citizen science, \O sterlund et al.~describe co-augmentation between volunteers and a machine-learning classifier: the classifier helps volunteers learn, and their labels retrain it \cite{refMutual}. Gaggioli et al.~call a related mode of human--AI creativity synergy: human and AI contribute in dialogue and shape each other's contributions \cite{refGaggioli}.

Both accounts treat human and machine as coupled contributors rather than user and tool, a view with roots in distributed cognition \cite{ref11} and the extended mind \cite{refExtended}. More recently, Clark examines how generative AI can take part in cognitive extension \cite{refClark}, Collins et al.~develop the notion of AI thought partners \cite{ref12}, and Gonzalez et al.~organize human--AI complementarity around reasoning, memory, and attention \cite{refGonzalez}. Unlike the classifier in citizen science, the model considered here keeps its learned weights fixed: the person may learn durably, while the model's contributions change with the context carried forward.

Evidence on whether human--AI collaboration helps is mixed. A meta-analysis across diverse tasks found that, on average, teams did better than humans alone but worse than AI alone when the AI was the stronger partner \cite{ref14}. In a word-guessing game, mixed human--AI groups outperformed human-only and AI-only groups when each player received the best previous guess \cite{refCreativeSearch}. In a short-story experiment, AI-generated ideas made individual stories more creative but the stories more similar to one another \cite{refDoshiHauser}. These studies measure outcomes: they show when collaboration helps, but not exactly how.

The word-guessing study makes one part of this process explicit: the best previous guess becomes the starting point for the next contribution. In dialogue, what passes from one exchange to the next is the person's reply together with the context the model retains, and together they shape the activity behind the model's next response (Figure~\ref{fig:loop}). Choosing which context to retain is the role of the spine.

\begin{figure}[!t]
\centering
\resizebox{\linewidth}{!}{%
\begin{tikzpicture}[x=0.01cm,y=-0.01cm]
\definecolor{figink}{HTML}{243846}
\definecolor{figmuted}{HTML}{536B78}
\definecolor{figteal}{HTML}{187D83}
\definecolor{figgold}{HTML}{B77B34}
\definecolor{figpale}{HTML}{E6ECEF}
\definecolor{figpanel}{HTML}{F3F7F8}
\definecolor{figlight}{HTML}{F0F7F6}
\definecolor{figwhite}{HTML}{FFFFFF}
\draw[draw=figteal,line width=0.91pt,line join=round,-{Stealth[length=2.1mm]}] (120,185) -- (120,85) -- (1340,85) -- (1340,185);
\node[anchor=base,inner sep=0pt,text=figteal,font=\sffamily\bfseries\fontsize{9.35}{11.55}\selectfont] at (730,49) {The ``spine'' selects context to carry forward};
\node[anchor=base,inner sep=0pt,text=figmuted,font=\sffamily\bfseries\fontsize{7.65}{9.45}\selectfont] at (320,146) {EARLIER EXCHANGE};
\node[anchor=base,inner sep=0pt,text=figmuted,font=\sffamily\bfseries\fontsize{7.65}{9.45}\selectfont] at (1140,146) {LATER EXCHANGE};
\path[fill=figlight,rounded corners=4.25pt] (90,185) rectangle (550,290);
\node[anchor=base,inner sep=0pt,text=figteal,font=\sffamily\bfseries\fontsize{8.79}{10.85}\selectfont] at (320,226) {Context};
\node[anchor=base,inner sep=0pt,text=figink,font=\sffamily\fontsize{7.37}{9.10}\selectfont] at (320,268) {A question or unresolved connection};
\path[fill=figlight,rounded corners=4.25pt] (910,185) rectangle (1370,290);
\node[anchor=base,inner sep=0pt,text=figteal,font=\sffamily\bfseries\fontsize{8.79}{10.85}\selectfont] at (1140,226) {New question};
\node[anchor=base,inner sep=0pt,text=figink,font=\sffamily\fontsize{7.94}{9.80}\selectfont] at (1140,268) {+ selected context};
\draw[draw=figmuted,line width=0.79pt,line join=round,-{Stealth[length=2.1mm]}] (320,300) -- (320,355);
\draw[draw=figmuted,line width=0.79pt,line join=round,-{Stealth[length=2.1mm]}] (1140,300) -- (1140,355);
\path[fill=figpanel,rounded corners=4.25pt] (90,365) rectangle (550,560);
\draw[draw=figpale,line width=0.57pt,line join=round] (190,388) -- (320,388);
\draw[draw=figpale,line width=0.57pt,line join=round] (190,388) -- (320,430);
\draw[draw=figpale,line width=0.57pt,line join=round] (190,388) -- (320,472);
\draw[draw=figpale,line width=0.57pt,line join=round] (190,430) -- (320,388);
\draw[draw=figpale,line width=0.57pt,line join=round] (190,430) -- (320,430);
\draw[draw=figpale,line width=0.57pt,line join=round] (190,430) -- (320,472);
\draw[draw=figpale,line width=0.57pt,line join=round] (190,472) -- (320,388);
\draw[draw=figpale,line width=0.57pt,line join=round] (190,472) -- (320,430);
\draw[draw=figpale,line width=0.57pt,line join=round] (190,472) -- (320,472);
\draw[draw=figpale,line width=0.57pt,line join=round] (320,388) -- (450,388);
\draw[draw=figpale,line width=0.57pt,line join=round] (320,388) -- (450,430);
\draw[draw=figpale,line width=0.57pt,line join=round] (320,388) -- (450,472);
\draw[draw=figpale,line width=0.57pt,line join=round] (320,430) -- (450,388);
\draw[draw=figpale,line width=0.57pt,line join=round] (320,430) -- (450,430);
\draw[draw=figpale,line width=0.57pt,line join=round] (320,430) -- (450,472);
\draw[draw=figpale,line width=0.57pt,line join=round] (320,472) -- (450,388);
\draw[draw=figpale,line width=0.57pt,line join=round] (320,472) -- (450,430);
\draw[draw=figpale,line width=0.57pt,line join=round] (320,472) -- (450,472);
\draw[draw=figteal,line width=1.13pt,line join=round] (190,388) -- (320,388);
\draw[draw=figteal,line width=1.13pt,line join=round] (320,388) -- (450,430);
\draw[draw=figteal,line width=1.13pt,line join=round] (190,430) -- (320,430);
\draw[draw=figteal,line width=1.13pt,line join=round] (320,430) -- (450,472);
\fill[figteal] (190,388) circle[radius=0.08cm];
\fill[figteal] (190,430) circle[radius=0.08cm];
\fill[figpale] (190,472) circle[radius=0.08cm];
\fill[figteal] (320,388) circle[radius=0.08cm];
\fill[figteal] (320,430) circle[radius=0.08cm];
\fill[figpale] (320,472) circle[radius=0.08cm];
\fill[figpale] (450,388) circle[radius=0.08cm];
\fill[figteal] (450,430) circle[radius=0.08cm];
\fill[figteal] (450,472) circle[radius=0.08cm];
\node[anchor=base,inner sep=0pt,text=figink,font=\sffamily\bfseries\fontsize{8.22}{10.15}\selectfont] at (320,530) {Patterns of model activity};
\path[fill=figpanel,rounded corners=4.25pt] (910,365) rectangle (1370,560);
\draw[draw=figpale,line width=0.57pt,line join=round] (1010,388) -- (1140,388);
\draw[draw=figpale,line width=0.57pt,line join=round] (1010,388) -- (1140,430);
\draw[draw=figpale,line width=0.57pt,line join=round] (1010,388) -- (1140,472);
\draw[draw=figpale,line width=0.57pt,line join=round] (1010,430) -- (1140,388);
\draw[draw=figpale,line width=0.57pt,line join=round] (1010,430) -- (1140,430);
\draw[draw=figpale,line width=0.57pt,line join=round] (1010,430) -- (1140,472);
\draw[draw=figpale,line width=0.57pt,line join=round] (1010,472) -- (1140,388);
\draw[draw=figpale,line width=0.57pt,line join=round] (1010,472) -- (1140,430);
\draw[draw=figpale,line width=0.57pt,line join=round] (1010,472) -- (1140,472);
\draw[draw=figpale,line width=0.57pt,line join=round] (1140,388) -- (1270,388);
\draw[draw=figpale,line width=0.57pt,line join=round] (1140,388) -- (1270,430);
\draw[draw=figpale,line width=0.57pt,line join=round] (1140,388) -- (1270,472);
\draw[draw=figpale,line width=0.57pt,line join=round] (1140,430) -- (1270,388);
\draw[draw=figpale,line width=0.57pt,line join=round] (1140,430) -- (1270,430);
\draw[draw=figpale,line width=0.57pt,line join=round] (1140,430) -- (1270,472);
\draw[draw=figpale,line width=0.57pt,line join=round] (1140,472) -- (1270,388);
\draw[draw=figpale,line width=0.57pt,line join=round] (1140,472) -- (1270,430);
\draw[draw=figpale,line width=0.57pt,line join=round] (1140,472) -- (1270,472);
\draw[draw=figteal,line width=1.13pt,line join=round] (1010,388) -- (1140,388);
\draw[draw=figteal,line width=1.13pt,line join=round] (1140,388) -- (1270,430);
\draw[draw=figgold,line width=1.13pt,line join=round] (1010,472) -- (1140,430);
\draw[draw=figgold,line width=1.13pt,line join=round] (1140,430) -- (1270,388);
\fill[figteal] (1010,388) circle[radius=0.08cm];
\fill[figpale] (1010,430) circle[radius=0.08cm];
\fill[figteal] (1010,472) circle[radius=0.08cm];
\fill[figteal] (1140,388) circle[radius=0.08cm];
\fill[figteal] (1140,430) circle[radius=0.08cm];
\fill[figpale] (1140,472) circle[radius=0.08cm];
\fill[figteal] (1270,388) circle[radius=0.08cm];
\fill[figteal] (1270,430) circle[radius=0.08cm];
\fill[figpale] (1270,472) circle[radius=0.08cm];
\node[anchor=base,inner sep=0pt,text=figink,font=\sffamily\bfseries\fontsize{8.22}{10.15}\selectfont] at (1140,530) {Activity recruited afresh};
\draw[draw=figmuted,line width=0.85pt,line join=round,dashed,-{Stealth[length=2.1mm]}] (320,570) -- (320,623);
\draw[draw=figmuted,line width=0.85pt,line join=round,dashed,-{Stealth[length=2.1mm]}] (1140,570) -- (1140,623);
\path[fill=figlight,rounded corners=4.25pt] (90,633) rectangle (550,733);
\node[anchor=base,inner sep=0pt,text=figteal,font=\sffamily\bfseries\fontsize{9.07}{11.20}\selectfont] at (320,674) {A new connection};
\node[anchor=base,inner sep=0pt,text=figink,font=\sffamily\fontsize{7.65}{9.45}\selectfont] at (320,710) {A possible change of perspective};
\path[fill=figlight,rounded corners=4.25pt] (910,633) rectangle (1370,733);
\node[anchor=base,inner sep=0pt,text=figteal,font=\sffamily\bfseries\fontsize{7.94}{9.80}\selectfont] at (1140,674) {A further connection};
\node[anchor=base,inner sep=0pt,text=figink,font=\sffamily\fontsize{7.37}{9.10}\selectfont] at (1140,710) {Reframing or extending the inquiry};
\path[fill=figpanel,rounded corners=4.25pt] (610,591) rectangle (850,733);
\node[anchor=base,inner sep=0pt,text=figink,font=\sffamily\bfseries\fontsize{7.94}{9.80}\selectfont] at (730,628) {Human response};
\node[anchor=base,inner sep=0pt,text=figink,font=\sffamily\fontsize{6.52}{8.05}\selectfont] at (730,667) {Further observations};
\node[anchor=base,inner sep=0pt,text=figink,font=\sffamily\fontsize{7.37}{9.10}\selectfont] at (730,704) {A new question};
\draw[draw=figmuted,line width=0.79pt,line join=round,-{Stealth[length=2.1mm]}] (560,681) -- (600,681);
\draw[draw=figmuted,line width=0.79pt,line join=round,-{Stealth[length=2.1mm]}] (730,581) -- (730,240) -- (900,240);
\node[anchor=base,inner sep=0pt,text=figmuted,font=\sffamily\fontsize{7.09}{8.75}\selectfont] at (625,384) {Same model};
\node[anchor=base,inner sep=0pt,text=figmuted,font=\sffamily\fontsize{7.09}{8.75}\selectfont] at (625,419) {Fixed weights};
\end{tikzpicture}
}
\caption{\textbf{Context selection and model activity in insightful dialogue.}
A model response can help a person connect further observations; the resulting interpretation returns through the next prompt. The proposed role of the ``spine'' is to carry forward context that helps elicit activity supporting further connections. The two network diagrams illustrate partly overlapping patterns of activity in the same model across successive exchanges, with learned weights held fixed. In the later network, teal paths indicate shared activity and gold paths newly recruited activity. The patterns are schematic and may involve distributed representations across shared components. Dashed arrows mark the proposed relation between model activity and the model's contributions.}
\label{fig:loop}
\par\smallskip{\footnotesize\raggedright\textbf{Alt text:} Two successive exchanges with the same model. An earlier response prompts further human observations and a new question. Selected context is carried forward to the later exchange, where partly overlapping patterns of model activity could support a further connection. Dashed arrows identify the hypothesized activity--contribution link.\par}
\end{figure}

\begin{samepage}
Three questions follow:

\textbf{Is there a signature of insight in model activity?} When a person's next message expresses insight, did the model's activity change during the preceding response in a way that recurs across topics?

\textbf{Can carried-forward context make the model enhance rather than parrot?} Which parts of an earlier exchange bring back activity associated with insight, without pulling the model back to the same interpretation?

\textbf{How do the two recurrences interact?} Can this coupled process be captured by a low-dimensional dynamical model?

\end{samepage}

The first two questions can be studied starting with records that already exist. Public collections contain real human--model dialogues \cite{refWildChat,refLMSYS}; for exchanges generated by an open-weight model, replaying the recorded tokens can approximate the model's activity along the conversation. The first step is to look for activity patterns that recur across topics before a person expresses insight. Adding these patterns to a predictor based on the preceding text would test whether they help anticipate such replies. The replies can indicate a change in understanding, but not establish when it happened or whether it was sudden. Carried-forward context can then be tested by intervention: changing what is retained, or steering candidate activity with existing methods \cite{refRepE,refCAA}, and measuring whether new connections appear in subsequent model contributions. The third question calls for a dynamical description, taken up in the next section. Throughout, a connection counts only if it is new and useful \cite{refCreativityDefinition} and moves the inquiry forward.

\subsection{A Neuroscience Perspective}

Neuroscience relates recorded neural activity to behavior. Dialogue with CK permits the same approach, with an advantage: in an open-weight model, the activity of every unit can be recorded and perturbed while the conversation unfolds \cite{ref17,ref18,refVyas}. The field's techniques therefore apply directly. Recorded activity can be reduced to a few latent variables whose dynamics are driven by inputs, in this case the person's messages \cite{refVyas}. It can be aligned to moments in the conversation, as neural recordings are aligned to a stimulus or a decision, to reveal what happens before a person expresses insight. And it can be perturbed, by steering or by changing the context carried forward, to test whether a pattern of activity contributes to subsequent model responses.

Information theory can help characterize how model activity relates to later insight, and when and where relevant patterns appear. In neuroscience, such measures show how early, and in which areas, neural activity carries information about an upcoming decision. Applied to dialogue, they could help identify how early in a response, and in which layers, activity is associated with a person's subsequent insight. Transfer entropy adds direction in time: it measures how much the past of one signal reduces uncertainty about the next value of another, beyond what the other's own past provides \cite{refTransfer}; Goodall et al.~develop related measures for language \cite{refGoodall}. Here it would ask whether the model's activity during a response tells more about the person's next reply than the person's earlier replies already do.

\subsection{Final Thoughts}

CK is a chorus of humanity, distilled into probabilistic form, capable of mimicry and of new synthesis. The proposal here is that part of the intelligence expressed in dialogue is realized through two interacting recurrences: generation conditioned on its own output, and the exchange through a person whose understanding can change. The modes describe how the model contributes; the spine decides what is carried forward. Insight happens in the person, but the model's contribution to it may have a signature: a transition in activity during the response preceding insight. Neuroscience offers the tools to look for it. If it can be found, and the context that recruits it carried forward, dialogue could be shaped to make insight more likely.

\subsection{Acknowledgments}

The reflections in this article draw on interactions with ChatGPT (GPT-4 and GPT-5). ChatGPT and Codex (GPT-6), provided by OpenAI, were used for assistance with structuring, language editing, literature searches, critical discussion, and drafting revisions. Claude Opus 5.5, provided by Anthropic, assisted with editing and fact-checking. The author is responsible for the article's arguments and final text.

\subsection{Funding}

The author received no specific funding for this work.

\subsection{Competing Interests}

The author declares no competing interests.

\subsection{Data Availability}

This Perspective reports no new datasets or analysis code. The sources supporting the discussion are cited in the text.

\end{document}